\def\tj{$t$-$J$}
\def\tjv{$t$-$J$-$V$}
\def\CITE{\onlinecite}

\documentclass[aps,prl,amsfonts,
twocolumn%
]{revtex4}
\usepackage{times}
\usepackage{epsf}
\usepackage{bm}
\usepackage{url}
\usepackage[hypertex]{hyperref}

\begin{document}

\title{Incipient order in the $\bm t$-$\bm J$ model at high temperatures}

\author{Leonid P. Pryadko}
\email[]{leonid@landau.ucr.edu}
\affiliation{Department of Physics, University of California,
  Riverside, California  92521} 
\author{Steven A. Kivelson}
\affiliation{Department of Physics, University of California, Los
  Angeles, California  90095}
\author{Oron Zachar}
\affiliation{Department of Physics, University of California, Los
  Angeles, California  90095}

\date{June 12, 2003}

\begin{abstract}
  We analyze the high-temperature behavior of the susceptibilities
  towards a number of possible ordered states in the \tjv\ model using
  the high-temperature series expansion.  From all diagrams with up to
  ten edges, reliable results are obtained down to temperatures of
  order $J$, or (with some optimism) to $J/2$.  In the unphysical
  regime, $t<J$, large superconducting susceptibilities are found,
  which moreover increase with decreasing temperatures, but for $t>J$,
  these susceptibilities are small and decreasing with decreasing
  temperature; this suggests that the \tj\ model does not support
  high-temperature superconductivity.  We also find modest evidence of
  a tendency toward nematic and $d$-density wave orders.
\end{abstract}

\maketitle

The discovery of high temperature (high-$T_c$) superconductivity in
the cuprate perovskites launched a renewed effort to develop an
understanding of the physics of highly correlated electronic systems.
It is clearly significant that superconductivity arises in these
materials upon doping a nearly ideal, spin 1/2 antiferromagnetic
insulating ``parent'' state.  Indeed, there is a prominent school of
thought\cite{Anderson-1987} that holds that high-$T_c$
superconductivity is more or less inevitable in a doped quasi two
dimensional (2D) antiferromagnet; for this reason enormous effort
has been focused on studies of the \tj\ model [Eq.~(\ref{eq:tjv})],
as the simplest model of a doped antiferromagnet.  However, while the
existence of antiferromagnetic order in the spin 1/2 Heisenberg model
(the zero doping limit of the \tj\ model) is well
established\cite{Chakravarty-88}, it remains uncertain
whether the 2D \tj\ model, by itself, supports high-$T_c$
superconductivity.  In addition to antiferromagnetism and
superconductivity, various other types of
order\cite{%
  Affleck-Marston-1988,
  Tranquada-95,kivelson-rmp-endnote,%
  Mook-Dai-Dogan-2001,
  Ando-nematic-2002} have been or may have been observed in the
cuprates, including nematic\cite{Ando-nematic-2002} (or spontaneous
breaking of the point-group symmetry),
charge-stripe,\cite{Tranquada-95} spin-stripe\cite{Tranquada-95}
and $d$-density\cite{Mook-Dai-Dogan-2001} wave
(also called staggered flux or orbital antiferromagnetic) order.  It
is thus interesting to determine which if any of these orders are
generic features of a doped antiferromagnet.

In this paper we report the results of an extensive high-temperature
series (HTS) study of the susceptibilities of the 2D \tj\ model toward
various short-period orders.  (Long-period stripe order cannot be
readily studied using these methods.)  Naturally, the results obtained
in this way are only reliable at moderately high temperatures.
However, corresponding to any low-temperature broken symmetry state
there must be a susceptibility which diverges at the ordering
transition; unless the transition is strongly first order, this
susceptibility will be large, and an increasing function of decreasing
temperature even at temperatures well above any ordering transition.

Put another way, the HTS is sensitive to relatively short-distance
physics (the range is determined by the order to which the series is
computed).  However, since the superconducting coherence length in the
cuprates is thought to be around two lattice constants, it is
reasonable to expect that the $10$-$12$ terms we have computed in this
series are sufficient to probe the physics of the model on
length-scales relevant to superconductivity.  It is important to
stress that HTS\cite{Domb-Green-book} is free of finite
size effects that plague other computational techniques.

Specifically, we have computed the HTS for antiferromagnetic (AF),
$d$-wave superconducting ($d$-SC), extended $s$-wave superconducting
($s$-SC), nematic (N), and orbital antiferromagnetic ($d$-DW)
susceptibilities, as defined in Eq.~(\ref{eq:chis}).  To extend the
temperature range over which the results can be trusted, we have made
use of standard methods of
resummation,\cite{Singh-1992-One} which we describe
explicitly below.  In outline, what we do is to construct a set of
Pad\'e approximants of related Euler-transformed series [see
Eq.~(\ref{eq:euler-transformation})] with different values of the parameter
$\beta_0$, eliminate all ``defective'' members of this set, and then
average over the remaining series. The variance in this average gives
an estimate for associated errors; where the variance gets large we
conclude that the results can no longer be trusted.  It is possible
that by biasing the series, using additional information about the low
temperature state obtained from other methods, one might be able to
extend the results to lower temperatures.  However, without such
additional information, it is our experience that using different
prescriptions for resummation, or even adding a few additional terms
to the series, does not significantly change the results or increase
their range of validity.

There have been previous high quality series
studies\cite{Kubo-Tada-1983-one,
  Singh-1992-One,Putikka-1992,%
  tenHaaf-1995,Putikka-1998} of this same model which
inspired the present work, but they primarily focused on extrapolating
the results to $T=0$.  What distinguishes the present study from these
earlier studies is (a) we have obtained susceptibilities that were not
previously computed\cite{Singh-1992-One,sokol} and (b) we have
contented ourselves 
with studying 
the high temperature behavior of the model.

\begin{figure}
   \centering
   \epsfxsize=3.4in
   \epsfbox{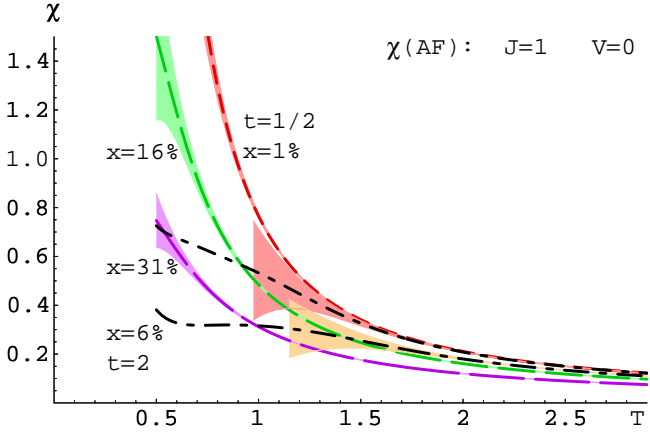}
   \caption{Temperature dependence of the AF susceptibility for
     $t=J/2$ (dashes, $x=1\%$, $16\%$, $31\%$) and $t=2J$ (dot-dashed
     lines, $x=1\%$, $6\%$) from the HTS to $1/T^{11}$ (diagrams
     with $N_E\le10$).  Shading represents the standard deviation
     of non-defective Pad\'e approximants as discussed in text.}
   \label{fig:AF}
 \end{figure}

\begin{figure}[htbp]
  \centering
  \epsfxsize=3.4in
  \epsfbox{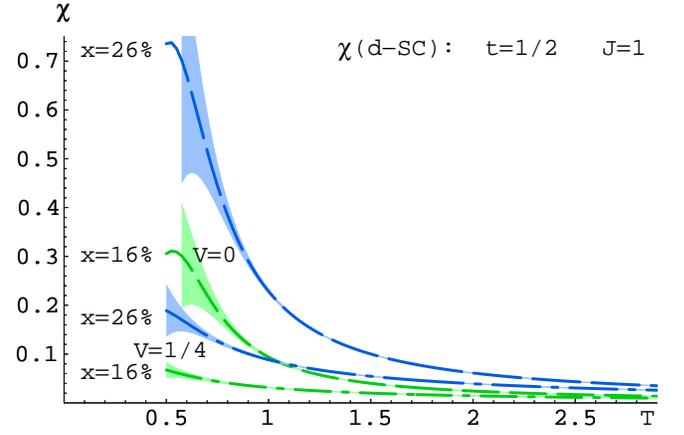}
  \caption{Superconducting ($d$-wave) susceptibility $\chi_{\rm d-SC}$
    for $t=J/2$ with $V=0$ (dashes) and $V=J/4$ (dot-dash) from the
   HTS to $1/T^{9}$ (diagrams with $N_E\le10$).  Shading
    as in Fig.~\protect\ref{fig:AF}.  Pairing fluctuations are strong
    at $V=0$, but are strongly suppressed already at $V=J/4$.}
  \label{fig:dscH}
\end{figure}

\begin{figure}[htbp]
  \centering
  \epsfxsize=3.4in
  \epsfbox{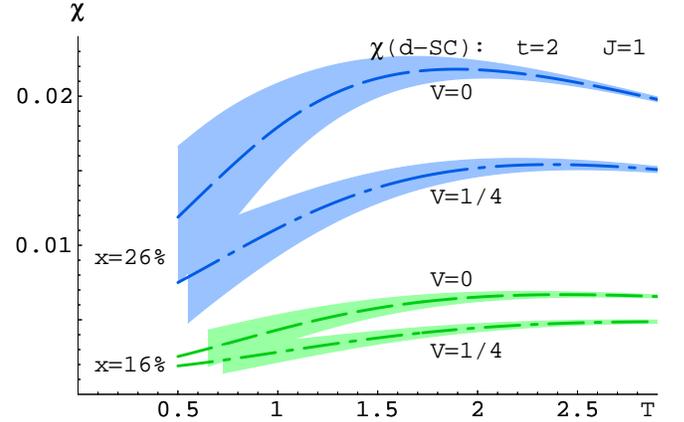}
  \caption{Same as Fig.~\protect\ref{fig:dscH} 
    for $t=2J$.  Pairing fluctuations are weak already at $V=0$
    (dashes), and decrease further with introduction of weak n.n.\ 
    repulsion (dot-dash).}
  \label{fig:dsc2}
\end{figure}

\begin{figure}[htbp]
  \centering
  \epsfxsize=3.4in
  \epsfbox{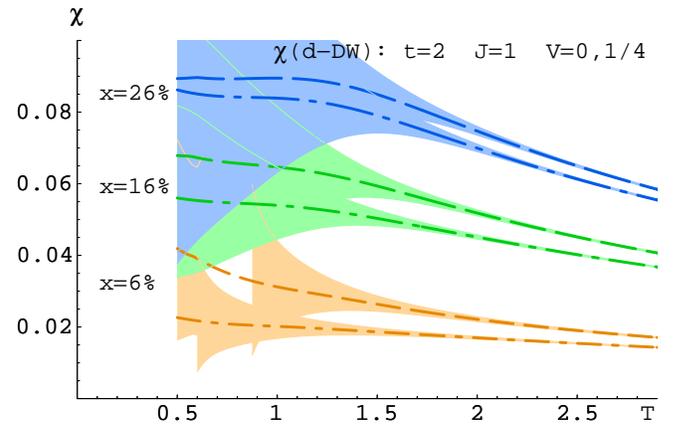}
  \caption{Staggered-flux susceptibility $\chi_{d\rm-DW}$
    for $t=2J$, at $V=0$ (dashes) and $V=J/4$ (dot-dash) obtained from
    HTS to $1/T^{9}$ ($N_E\le10$).   
    Shading as in Fig.~\protect\ref{fig:AF}.}
  \label{fig:ddw2}
\end{figure}

\begin{figure}[htbp]
  \centering
  \epsfxsize=3.4in
  \epsfbox{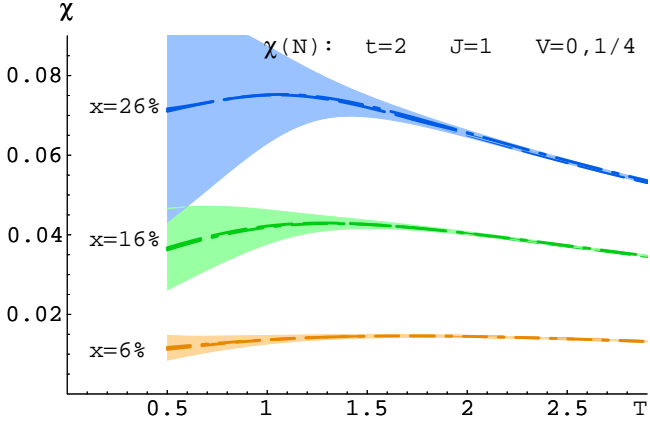}
  \caption{Temperature dependence of the nematic susceptibility $\chi_{\rm N}$
    for $t=2J$ and $V=0$ (dashes) and $V=J/4$ (dot-dash) obtained from
    HTS to $1/T^{9}$ ($N_E\le10$). Shading as in Fig.~\protect\ref{fig:AF}.
    The n.n.\ repulsion term $V$ has virtually no effect on
    $\chi_{\rm N}$.}
  \label{fig:n2}
\end{figure}

Our principal findings are as follows: {\bf 1}.~The results we obtain
are reliable, without apology, for temperatures $T > J$, and are
probably qualitatively correct down to $T\sim J/2$.  However, we have
not found any method of analyzing the series that we trust to any
lower temperatures.  {\bf 2}.~In the ``unphysical'' range of
parameters, $t\le J$, with $V=0$, the strongest incipient order is
AF---the AF susceptibility, $\chi_{\rm AF}$, is large and
shows a strong tendency to increase with decreasing temperature,
although this tendency gets gradually weaker with increasing doping
$x$ (Fig.~\ref{fig:AF}).  The SC susceptibilities are  largest at values of doping
$x\gtrsim 16\%$ (both $d$-SC shown in Fig.~\ref{fig:dscH} and $s$-SC,
not shown\cite{all-hts-nb}; typically
$\chi_{d\rm-SC}>\chi_{s\rm-SC}$).  However, the inclusion of an
additional nearest-neighbor (n.n.)  repulsion, $V = J/4$, is
sufficient to strongly suppress the pairing fluctuations.  {\bf 3}.~In the
physical range of parameters, $t > J$, where the bare interactions
between electrons are truly repulsive, both $\chi_{d\rm-SC}$
(Fig.~\ref{fig:dsc2}) and $\chi_{s\rm-SC}$ (not
shown\cite{all-hts-nb}) are small and decreasing with decreasing
temperature already for $T\lesssim 2J$; the pairing fluctuations are
further suppressed by the addition of small n.n.\ repulsion.  Based
on these observations we conclude that {\em the 2D \tj\ model with $t
  > J$ probably does not support high temperature
  superconductivity\/}.  {\bf 4}.~For $t>J$, commensurate AF
fluctuations are moderate for $x=1\%$ but are
dramatically suppressed already at $x\ge 6\%$ (Fig.~\ref{fig:AF}).  At
larger $x$, we find 
that the $d$-DW (Fig.~\ref{fig:ddw2}) and the nematic
(Fig.~\ref{fig:n2}) susceptibilities are both moderate, show a weak
tendency to increase with decreasing temperature, and remain virtually
unaffected by the addition of a weak n.n.\ repulsion, $V=J/4$.


We compute the high-temperature series for the 2D \tj\ model defined on
the square lattice, 
\begin{equation}
  \label{eq:tjv}
  H=-\sum_{\langle ij\rangle} t_{ij} \,c^\dagger_{i\sigma}c_{j\sigma}
  +J\,\Bigl({\bf S}_i\cdot {\bf S}_j-{1\over 4}n_in_j\Bigr)+V\, n_i n_j,
\end{equation}
where $t_{ij}=t$ is the hopping matrix element, $c_{i\sigma}$
is the 
electron annihilation operator, ${\bf S}_i\equiv
(1/2)c^\dagger_{i\sigma}{\bm\tau}_{\sigma\sigma'}c_{i\sigma'}$ and
$n_i\equiv c^\dagger_{i\sigma}c_{i\sigma}$ are on-site spin and charge
operators, and the doubly-occupied sites are projected out.  The
canonical \tj\ model\cite{Anderson-1987} can be obtained
from Eq.~(\ref{eq:tjv}) by setting the n.n.\ repulsion $V=0$, while the version of
the model used in previous
high-temperature series studies\cite{%
  Kubo-Tada-1983-one,
  Singh-1992-One,%
  Putikka-1992,Putikka-1998} can be obtained by setting
$V=J/4$.  All results presented in this paper refer to these two
values of $V$.

The susceptibilities (per unit site) $\chi$ can be expressed via the
second derivatives of the free energy with respect to appropriately
chosen perturbing parameters, or as the irreducible thermodynamic
correlation functions of (properly projected) operators $\mathcal{O}$,
\begin{equation}
  \label{eq:chis}
  \chi_{\mathcal{O}}\equiv {1\over N}\,T_\tau \!\int_0^\beta\!\! d\tau\,
  \bigl\langle\!\bigl\langle
  e^{H \tau}  \mathcal{O} e^{-H \tau}
  \mathcal{O}^\dagger\bigr\rangle\!\bigr\rangle_\beta,\quad
  \beta\equiv 1/T.
\end{equation}
We take the staggered magnetization $\mathcal{O}_{\rm AF}=\sum_{\bf r}
e^{i{\bf Q}\cdot{\bf r}} S^z_{\bf r}$, ${\bf Q}\equiv (\pi,\pi)$, for
AF ordering, the anisotropic part of the kinetic energy ${\cal O}_{\rm
  N}= \sum_{\bf r}\bigl(c^\dagger_{{\bf r}+\hat{\bf x}\,\sigma}c_{{\bf
    r}\,\sigma} - c^\dagger_{{\bf r}+\hat {\bf y}\,\sigma}c_{{\bf
    r}\,\sigma}+{\rm h.c.}\bigr)/2 $ for nematic ordering, the
staggered orbital currents ${\cal O}_{d{\rm-DW}}= i\sum_{\bf r}e^{i\bf Q\cdot
  r}\bigl( c^\dagger_{{\bf r}+\hat{\bf y}\,\sigma}c_{{\bf r}\,\sigma}-
c^\dagger_{{\bf r}+\hat{\bf x}\,\sigma}c_{{\bf r}\,\sigma}-{\rm
  h.c.}\bigr)/2 $ for $d$-DW ordering\cite{endnote-ddw}, and the
isotropic (anisotropic) part of the uniform pairing $ \mathcal{O}_{\rm
  SC}= \sum_{\bf r}\left(\Delta_{{\bf r},{\bf r}+\hat{\bf x}}\pm
  \Delta_{{\bf r},{\bf r}+\hat{\bf y}}\right)/2 $ for $s$-SC ($d$-SC)
ordering, where the pair annihilation operator $
\Delta_{ij}\equiv 
c_{i\uparrow} c_{j\downarrow} + c_{j\uparrow}c_{i\downarrow}
$.  Respectively, the first coefficients of the corresponding series
are 
$\chi^{(1)}_{\rm AF}=\beta (1-x)/4$, 
$\chi^{(1)}_{\rm N}=
\chi^{(1)}_{d\rm -DW}=\beta x(1-x)$, 
and 
$\chi^{(1)}_{d\rm -SC}=
\chi^{(1)}_{s\rm -SC}=\beta x^2$. 

The definition of susceptibilities in terms of the derivatives of the
free energy offers a convenient way for constructing the cluster
expansion\cite{Domb-Green-book}.  For each inequivalent
lattice placement ${\cal C}$ of a given diagram (connected graph) with
$n_E$ edges, the relevant traces (grouped by the number of particles)
are computed using block-diagonal matrices of the cluster Hamiltonian
and the operator ${\cal O}$.  The traces are combined to produce the
coefficients of the inverse-temperature expansion of the cluster
susceptibility $\chi_{\cal O}(\beta,y; {\cal C})$ and the
thermodynamic potential $\Omega(\beta,y;{\cal C})$ dependent on the
variable $y=z/(1+2z)$ related to fugacity $z=e^{\beta\mu}$.  After the
subcluster subtraction we obtain the irreducible weights of the
cluster [whose expansion starts with ${\cal O}(\beta^{n_E-1})$ or
higher power of $\beta$].  Then,
combining the cluster weights for diagrams with up to $N_E$ edges, we
generate the series exact in the thermodynamic limit to  $\sim\beta^{N_E-1}$
($\beta^{N_E+1}$ for AF order since the operator ${\cal O}_{{\rm AF}}$
is defined on the vertices).  As the last step, we perform the
Legendre transformation to obtain the series for the free energy
$F(x,\beta)$, and reexpress\cite{Kubo-Tada-1983-one}(b) the obtained
series for $\chi$ in terms of the average hole density
$x=1+(\partial\Omega/\partial\mu)_T$.

All obtained series for $\chi$, $\Omega$, and $F$ were carefully
compared with series computed analytically to $\beta^3$, and also with
series to $\beta^5$ obtained by a direct differentiation of the free
energy expression generated with an independently-written
Mathematica\cite{mathematica} program.  Among other consistency
checks, we verified the cancellation of low-order terms in the
irreducible weights.  We have also compared the obtained free energy
and related specific heat series with those for the \tj\ model with
$V=1/4$ from Refs.~\CITE{Singh-1992-One,Putikka-1992},
and also the specific heat series at $x=0$ with the corresponding
series for the Heisenberg model\cite{hts-heisenberg}.

A standard way of extrapolating a power series in $\beta$ is to
construct a ratio of polynomials $p_n(\beta)/q_m(\beta)$ with matching
expansion in powers of $\beta$, referred to as an $(n,m)$ Pad\'e
approximant.  However, when only a few first terms of the series are
known, and without a detailed knowledge of the structure of the
singularities of the function, the procedure is plagued by spurious
divergences.  Indeed, the coefficients of the power series are only
weakly modified if the numerator and the denominator of the fraction
have close roots.  The accuracy of the extrapolation would not be
affected by cancellation of such factors, which amounts to using
smaller $n$ and $m$.  Furthermore, it is not generally clear 
whether a series in $\beta$ would give better extrapolation than a
series in a related variable $\tilde\beta=f(\beta)$.  As a result, it
has become standard practice to
average\cite{Singh-1992-One} over a large number of
``non-defective'' Pad\'e approximants generated for a family of
functions $f(\beta,\beta_0)$ with different $\beta_0$, and use the
corresponding dispersion to estimate the errors.  A specific
difficulty of extrapolating the series for susceptibilities [compared
to non-singular quantities such as $n({\bf k})$] is that interesting
susceptibilities can actually diverge at a critical temperature.
Additionally, the peaks in $\chi$ develop at relatively large spatial
scale, meaning that they become pronounced only at sufficiently high
orders.

To generate the curves in Figs.~1--5, for each $\chi(\beta)$ at a
given set of parameters we constructed a large number of Pad\'e
approximants (all $n$, $m$ with $n+m\ge (n+m)_{\rm max}-4$) for both
the original series and a number of series in terms of the 
Euler-transformed variable\cite{troyer-hts}
\begin{equation}
  \label{eq:euler-transformation}
  \tilde\beta=\beta/(\beta_0+\beta),
\end{equation}
using $0.3\le\beta_0\le 10$.  We then eliminated as ``defective'' the
approximants with positive real roots in denominators and numerators
[susceptibilities are nonnegative, see Eq.~(\ref{eq:chis}), and we do
not expect an actual phase transition with a divergent $\chi$ at such
high temperatures], as well as the approximants with close
numerator-denominator complex root pairs, and calculated the averages and the
corresponding rms deviations for different temperatures $T=1/\beta$.

As an independent method of analysis, we have also looked at the
behavior of the untransformed and most closely balanced ($m$ close to
$n$) Pad\'e approximants with even $m$.  When they are non-defective,
these ``best'' approximants produce curves that look qualitatively
similar to the average, and lie well within the shaded regions of the
plots.  We have also tried a number of variations on the averaging
procedure, testing on the AF susceptibility $\chi_{\rm AF}$ at
$x=1\%$, where we expect a prominent peak at low temperatures.  In
particular, we found that the hyperbolic tangent transformation
$\tilde\beta=\tanh(\beta/\beta_0)$ which was popular in previous HTS
studies\cite{Singh-1992-One,Putikka-1992}, strongly suppresses all
peaks, effectively flattening the extrapolated functions for all
susceptibilities we computed.

To summarize, our results indicate that, although in the
``unphysical'' region, $t\lesssim J$, the \tj\ model displays a sharp
increase in the pairing fluctuations as the temperature goes down,
this is no longer the case for $t>J$.  Also, superconducting
fluctuations are strongly suppressed by introduction of a small n.n.\ 
repulsion $V$.  Thus we conclude that high-temperature
superconductivity is probably not a generic feature\cite{scalapino} of doped AFs.
Apart from $\chi_{\rm AF}$ at very small doping, none of the studied
susceptibilities display remarkably strong fluctuations in the
``physical'' range $t>J$ as the temperature goes down to $T\gtrsim
J/2$, but $\chi_{\rm N}$ and $\chi_{d\rm-DW}$ do show a modest
enhancement.

In the future, we intend to continue studying the \tjv\ model within
the high-temperature series approach, for a wider range of orders in
hope of identifying the most relevant fluctuations in the pseudogap
region.  We also intend to study the interaction of order parameters,
by looking at various susceptibilities in modified models where an
ordering (e.g., hopping anisotropy) is imposed at the Hamiltonian
level.  Within this general approach, we also plan to study a range of
related models, in particular an array of coupled \tj\ ladders.

{\bf Acknowledgments:} We are indebted to Sudip Chakravarty, Mattias
K\"orner, Rajiv Singh, and Matthias Troyer for enlightening
discussions.  This work was supported by the NSF grant DMR01-10329 (S.A.K.)
and the DOE grant DE-FG03-00ER45798 (O.Z.).  The
calculations were performed in part on the clustered computer
at the Institute of Geophysics and Planetary Physics, UC, Riverside.


\begin{thebibliography}{10}
\expandafter\ifx\csname bibnamefont\endcsname\relax
  \def\bibnamefont#1{#1}\fi
\expandafter\ifx\csname bibfnamefont\endcsname\relax
  \def\bibfnamefont#1{#1}\fi
\expandafter\ifx\csname url\endcsname\relax
  \def\url#1{\texttt{#1}}\fi
\expandafter\ifx\csname urlprefix\endcsname\relax\def\urlprefix{URL }\fi
\expandafter\ifx\csname bibinfo\endcsname\relax \def\bibinfo#1#2{#2}\fi
\expandafter\ifx\csname eprint\endcsname\relax \def\eprint#1{#1}\fi

\bibitem{Anderson-1987}
\bibinfo{author}{\bibfnamefont{P.~W.} \bibnamefont{Anderson}},
  \bibinfo{journal}{Science} \textbf{\bibinfo{volume}{235}},
  \bibinfo{pages}{1996} (\bibinfo{year}{1987});
  \emph{\bibinfo{title}{The theory of superconductivity in the high-Tc
  cuprates}} (\bibinfo{publisher}{Princeton Univ.\ Press},
  \bibinfo{address}{Princeton, NJ}, \bibinfo{year}{1997}).

\bibitem{Chakravarty-88}
\bibinfo{author}{\bibfnamefont{S.}~\bibnamefont{Chakravarty}},
  \bibinfo{author}{\bibfnamefont{B.~I.} \bibnamefont{Halperin}},
  \bibnamefont{and} \bibinfo{author}{\bibfnamefont{D.~R.}
  \bibnamefont{Nelson}}, \bibinfo{journal}{Phys. Rev. Lett.}
  \textbf{\bibinfo{volume}{60}}, \bibinfo{pages}{1057} (\bibinfo{year}{1988});
%
\bibinfo{journal}{Phys. Rev. B}
  \textbf{\bibinfo{volume}{39}}, \bibinfo{pages}{2344} (\bibinfo{year}{1989}).

\bibitem{Affleck-Marston-1988}
\bibinfo{author}{\bibfnamefont{I.}~\bibnamefont{Affleck}} \bibnamefont{and}
  \bibinfo{author}{\bibfnamefont{J.~B.} \bibnamefont{Marston}},
  \bibinfo{journal}{Phys. Rev. B} \textbf{\bibinfo{volume}{37}},
  \bibinfo{pages}{3774} (\bibinfo{year}{1988});
%
\bibinfo{author}{\bibfnamefont{J.}~\bibnamefont{Zaanen}} \bibnamefont{and}
  \bibinfo{author}{\bibfnamefont{O.}~\bibnamefont{Gunnarsson}},
  ibid., 
 \textbf{\bibinfo{volume}{40}},
  \bibinfo{pages}{7391} (\bibinfo{year}{1989});
%
\bibinfo{author}{\bibfnamefont{H.~J.} \bibnamefont{Schulz}},
  \bibinfo{journal}{Phys. Rev. Lett.} \textbf{\bibinfo{volume}{64}},
  \bibinfo{pages}{1445} (\bibinfo{year}{1990});
%
\bibinfo{author}{\bibfnamefont{K.}~\bibnamefont{Machida}},
  \bibinfo{journal}{Physica C: Superconductivity}
  \textbf{\bibinfo{volume}{158}}, \bibinfo{pages}{192} (\bibinfo{year}{1989});
%
\bibinfo{author}{\bibfnamefont{U.}~\bibnamefont{L{\"o}w}},
  \bibinfo{author}{\bibfnamefont{V.~J.} \bibnamefont{Emery}},
  \bibinfo{author}{\bibfnamefont{K.}~\bibnamefont{Fabricius}},
  \bibnamefont{and} \bibinfo{author}{\bibfnamefont{S.~A.}
  \bibnamefont{Kivelson}}, \bibinfo{journal}{Phys. Rev. Lett.}
  \textbf{\bibinfo{volume}{72}}, \bibinfo{pages}{1918} (\bibinfo{year}{1994});
\bibinfo{author}{\bibfnamefont{O.}~\bibnamefont{Zachar}},
  \bibinfo{author}{\bibfnamefont{S.~A.}~\bibnamefont{Kivelson}},
  \bibnamefont{and} 
  \bibinfo{author}{\bibfnamefont{V.~J.}~\bibnamefont{Emery}},
  \bibinfo{journal}{Phys. Rev. B} \textbf{\bibinfo{volume}{57}},
  \bibinfo{pages}{1422} (\bibinfo{year}{1998});
%
\bibinfo{author}{\bibfnamefont{S.-C.} \bibnamefont{Zhang}},
  \bibinfo{journal}{Science} \textbf{\bibinfo{volume}{275}},
  \bibinfo{pages}{1089} (\bibinfo{year}{1997});
%
\bibinfo{author}{\bibfnamefont{S.~A.} \bibnamefont{Kivelson}},
  \bibinfo{author}{\bibfnamefont{E.}~\bibnamefont{Fradkin}}, \bibnamefont{and}
  \bibinfo{author}{\bibfnamefont{V.~J.} \bibnamefont{Emery}},
  \bibinfo{journal}{Nature} \textbf{\bibinfo{volume}{393}},
  \bibinfo{pages}{550} (\bibinfo{year}{1998});
%
\bibinfo{author}{\bibfnamefont{S.}~\bibnamefont{Chakravarty}},
  \bibinfo{author}{\bibfnamefont{R.~B.} \bibnamefont{Laughlin}},
  \bibinfo{author}{\bibfnamefont{D.~K.} \bibnamefont{Morr}}, \bibnamefont{and}
  \bibinfo{author}{\bibfnamefont{C.}~\bibnamefont{Nayak}},
  \bibinfo{journal}{Phys. Rev. B} \textbf{\bibinfo{volume}{63}},
  \bibinfo{pages}{094503} (\bibinfo{year}{2001});
\bibinfo{author}{\bibfnamefont{S.}~\bibnamefont{Chakravarty}},
  \bibinfo{author}{\bibfnamefont{H.-Y.} \bibnamefont{Kee}}, \bibnamefont{and}
  \bibinfo{author}{\bibfnamefont{C.}~\bibnamefont{Nayak}},
  \bibinfo{journal}{Int. J. Mod. Phys. B} \textbf{\bibinfo{volume}{21}},
  \bibinfo{pages}{2901} (\bibinfo{year}{2001}).

\bibitem{Tranquada-95}
\bibinfo{author}{\bibfnamefont{J.~M.} \bibnamefont{Tranquada}},
  \bibinfo{author}{\bibfnamefont{B.~J.} \bibnamefont{Sternlieb}},
  \bibinfo{author}{\bibfnamefont{J.~D.} \bibnamefont{Axe}},
  \bibinfo{author}{\bibfnamefont{Y.}~\bibnamefont{Nakamura}}, \bibnamefont{and}
  \bibinfo{author}{\bibfnamefont{S.}~\bibnamefont{Uchida}},
  \bibinfo{journal}{Nature} \textbf{\bibinfo{volume}{375}},
  \bibinfo{pages}{561} (\bibinfo{year}{1995}).


\bibitem{kivelson-rmp-endnote}
\bibinfo{note}{For review of the status of experimental search for stripe and
  nematic order in the cuprates see
%
\bibinfo{author}{\bibfnamefont{S.~A.} \bibnamefont{Kivelson}},
  \bibinfo{author}{\bibfnamefont{E.}~\bibnamefont{Fradkin}},
  \bibinfo{author}{\bibfnamefont{V.}~\bibnamefont{Oganesyan}},
  \bibinfo{author}{\bibfnamefont{I.~P.} \bibnamefont{Bindloss}},
  \bibinfo{author}{\bibfnamefont{J.~M.} \bibnamefont{Tranquada}},
  \bibinfo{author}{\bibfnamefont{A.}~\bibnamefont{Kapitulnik}},
  \bibnamefont{and} \bibinfo{author}{\bibfnamefont{C.}~\bibnamefont{Howald}},
  \emph{\bibinfo{title}{How to detect fluctuating order in the high-temperature
  superconductors}}, \bibinfo{note}{cond-mat/0210683} [unpublished]}.


\bibitem{Mook-Dai-Dogan-2001}
\bibinfo{author}{\bibfnamefont{H.~A.} \bibnamefont{Mook}},
  \bibinfo{author}{\bibfnamefont{P.}~\bibnamefont{Dai}}, \bibnamefont{and}
  \bibinfo{author}{\bibfnamefont{F.}~\bibnamefont{Dogan}},
  \bibinfo{journal}{Phys. Rev. B} \textbf{\bibinfo{volume}{64}},
  \bibinfo{pages}{012502} (\bibinfo{year}{2001});
%
\bibinfo{author}{\bibfnamefont{H.~A.} \bibnamefont{Mook}},
  \bibinfo{author}{\bibfnamefont{P.}~\bibnamefont{Dai}},
  \bibinfo{author}{\bibfnamefont{S.~M.} \bibnamefont{Hayden}},
  \bibinfo{author}{\bibfnamefont{A.}~\bibnamefont{Hiess}},
  \bibinfo{author}{\bibfnamefont{J.~W.} \bibnamefont{Lynn}},
  \bibinfo{author}{\bibfnamefont{S.-H.} \bibnamefont{Lee}}, \bibnamefont{and}
  \bibinfo{author}{\bibfnamefont{F.}~\bibnamefont{Dogan}},
  ibid., 
  \textbf{\bibinfo{volume}{66}},
  \bibinfo{pages}{144513} (\bibinfo{year}{2002}).


\bibitem{Ando-nematic-2002}
\bibinfo{author}{\bibfnamefont{Y.}~\bibnamefont{Ando}},
  \bibinfo{author}{\bibfnamefont{K.}~\bibnamefont{Segawa}},
  \bibinfo{author}{\bibfnamefont{S.}~\bibnamefont{Komiya}}, \bibnamefont{and}
  \bibinfo{author}{\bibfnamefont{A.~N.} \bibnamefont{Lavrov}},
  \bibinfo{journal}{Phys. Rev. Lett.} \textbf{\bibinfo{volume}{88}},
  \bibinfo{pages}{137005} (\bibinfo{year}{2002}).

\bibitem{Domb-Green-book}
\bibinfo{editor}{\bibfnamefont{C.}~\bibnamefont{Domb}} \bibnamefont{and}
  \bibinfo{editor}{\bibfnamefont{M.~S.} \bibnamefont{Green}}, eds.,
  \emph{\bibinfo{title}{Phase transitions and critical phenomena}},
  vol.~\bibinfo{volume}{3} (\bibinfo{publisher}{Academic},
  \bibinfo{address}{London}, \bibinfo{year}{1974});
%
\bibinfo{author}{\bibfnamefont{J.}~\bibnamefont{G.~A.~Baker}},
  \emph{\bibinfo{title}{Quantitative theory of critical phenomena}}
  (\bibinfo{publisher}{Academic}, \bibinfo{address}{San Diego},
  \bibinfo{year}{1990}).

\bibitem{Singh-1992-One}
\bibinfo{author}{\bibfnamefont{R.~R.~P.} \bibnamefont{Singh}} \bibnamefont{and}
  \bibinfo{author}{\bibfnamefont{R.~L.} \bibnamefont{Glenister}},
  \bibinfo{journal}{Phys. Rev. B} \textbf{\bibinfo{volume}{46}},
  \bibinfo{pages}{11871} (\bibinfo{year}{1992});
%
  \textbf{\bibinfo{volume}{46}},
  \bibinfo{pages}{14313} (\bibinfo{year}{1992}).

\bibitem{Kubo-Tada-1983-one}
\bibinfo{author}{\bibfnamefont{K.}~\bibnamefont{Kubo}} \bibnamefont{and}
  \bibinfo{author}{\bibfnamefont{M.}~\bibnamefont{Tada}},
  \bibinfo{journal}{Progr. Theor. Phys. (Japan)} \textbf{\bibinfo{volume}{69}},
  \bibinfo{pages}{1345} (\bibinfo{year}{1983});
%
\textbf{\bibinfo{volume}{71}},
\bibinfo{pages}{479} (\bibinfo{year}{1984}).

\bibitem{Putikka-1992}
\bibinfo{author}{\bibfnamefont{W.~O.} \bibnamefont{Putikka}},
  \bibinfo{author}{\bibfnamefont{M.~U.} \bibnamefont{Luchini}},
  \bibnamefont{and} \bibinfo{author}{\bibfnamefont{T.~M.} \bibnamefont{Rice}},
  \bibinfo{journal}{Phys. Rev. Lett.} \textbf{\bibinfo{volume}{68}},
  \bibinfo{pages}{538} (\bibinfo{year}{1992}); 
\bibinfo{author}{\bibfnamefont{R.}~\bibnamefont{Hlubina}},
  \bibinfo{author}{\bibfnamefont{W.~O.} \bibnamefont{Putikka}},
  \bibinfo{author}{\bibfnamefont{T.~M.} \bibnamefont{Rice}}, \bibnamefont{and}
  \bibinfo{author}{\bibfnamefont{D.~V.} \bibnamefont{Khveshchenko}},
  \bibinfo{journal}{Phys. Rev. B} \textbf{\bibinfo{volume}{46}},
  \bibinfo{pages}{11224} (\bibinfo{year}{1992}).


\bibitem{tenHaaf-1995}
\bibinfo{author}{\bibfnamefont{D.~F.~B.} \bibnamefont{ten Haaf}},
  \bibinfo{author}{\bibfnamefont{P.~W.} \bibnamefont{Brouwer}},
  \bibinfo{author}{\bibfnamefont{P.~J.~H.} \bibnamefont{Denteneer}},
  \bibnamefont{and} \bibinfo{author}{\bibfnamefont{J.~M.~J.} \bibnamefont{van
  Leeuwen}}, \bibinfo{journal}{Phys. Rev. B} \textbf{\bibinfo{volume}{51}},
  \bibinfo{pages}{353} (\bibinfo{year}{1995}).

\bibitem{Putikka-1998}
\bibinfo{author}{\bibfnamefont{W.~O.} \bibnamefont{Putikka}},
  \bibinfo{author}{\bibfnamefont{M.~U.} \bibnamefont{Luchini}},
  \bibnamefont{and} \bibinfo{author}{\bibfnamefont{R.~R.~P.}
  \bibnamefont{Singh}}, \bibinfo{journal}{Phys. Rev. Lett.}
  \textbf{\bibinfo{volume}{81}}, \bibinfo{pages}{2966} (\bibinfo{year}{1998});
%
\bibinfo{author}{\bibfnamefont{W.~O.} \bibnamefont{Putikka}} \bibnamefont{and}
  \bibinfo{author}{\bibfnamefont{M.~U.} \bibnamefont{Luchini}},
  \bibinfo{journal}{Phys. Rev. B} \textbf{\bibinfo{volume}{62}},
  \bibinfo{pages}{1684} (\bibinfo{year}{2000}).
  
\bibitem{sokol} \bibinfo{note}{Note that $\chi_{\rm AF}$ was previously computed to
  the same order in Ref.~\protect\CITE{Singh-1992-One}(a) and in} A.
  Sokol, R. L. Glenister, and R. R. P. Singh, \bibinfo{journal}{Phys. Rev. Lett.}
  \textbf{\bibinfo{volume}{72}}, \bibinfo{pages}{1549}
  (\bibinfo{year}{1994}).

\bibitem{all-hts-nb}
  \bibinfo{note}{See additional plots
    at~\urlprefix\url{http://faculty.ucr.edu/~leonid/hts-all.pdf}.
    The calculated series are available from the authors upon
    request.}

\bibitem{endnote-ddw}
\bibinfo{note}{The gauge-invariant $d$-DW susceptibility defined as the second
  derivative of the free energy with respect to staggered fluxes has a
  qualitatively similar behavior for $x\gtrsim 11\%$ (where it is positive).}

\bibitem{mathematica}
\bibinfo{author}{\bibfnamefont{S.}~\bibnamefont{Wolfram}},
  \emph{\bibinfo{title}{Mathematica}}, \bibinfo{organization}{Wolfram Research,
  Inc}, \bibinfo{address}{Champaign, IL} (\bibinfo{year}{2001}).

\bibitem{hts-heisenberg}
\bibinfo{author}{\bibfnamefont{J.}~\bibnamefont{Oitmaa}} \bibnamefont{and}
  \bibinfo{author}{\bibfnamefont{E.}~\bibnamefont{Bornilla}},
  \bibinfo{journal}{Phys. Rev. B} \textbf{\bibinfo{volume}{53}},
  \bibinfo{pages}{14228} (\bibinfo{year}{1996}).

\bibitem{troyer-hts}
\bibinfo{author}{\bibfnamefont{M.}~\bibnamefont{Troyer}} \bibnamefont{and}
  \bibinfo{author}{\bibfnamefont{M.}~\bibnamefont{K{\"o}rner}}
  (\bibinfo{year}{2003}), \bibinfo{note}{private communication}.
  
\bibitem{scalapino} 
  \bibinfo{note}{We believe that the increase of $\chi_{d\rm-SC}$ with
  decreasing $T$ found by}  
\bibinfo{author}{\bibnamefont{S.} \bibnamefont{R.} \bibnamefont{White}},
\bibinfo{author}{\bibnamefont{D.} \bibnamefont{J.} \bibnamefont{Scalapino}}, 
\bibinfo{author}{\bibnamefont{R.} \bibnamefont{L.} \bibnamefont{Sugar}},
\bibinfo{author}{\bibnamefont{N.} \bibnamefont{E.}
  \bibnamefont{Bickers}}, and 
\bibinfo{author}{\bibnamefont{R.} \bibnamefont{T.} \bibnamefont{Scalettar}}, \bibinfo{journal}{Phys. Rev. B} 
\textbf{\bibinfo{volume}{39}}, \bibinfo{pages}{839}
  (\bibinfo{year}{1989}) 
  \bibinfo{note}{for the Hubbard model with $U=4t$ is not in conflict with our
  strong-coupling results.}
  
\end{thebibliography}
\end{document}


\title{Erratum: Incipient order in the $\bm t$-$\bm J$ model at high
  temperatures}

\author{Leonid P. Pryadko}
\affiliation{Department of Physics \& Astronomy, University of California,
  Riverside, California  92521} 
\author{Steven A. Kivelson}
\affiliation{Department of Physics, Stanford University, Stanford, California  94305}
\author{Oron Zachar}
 \affiliation{Department of Physics, University of California, Los
   Angeles, California  90095}

\date\today

\begin{abstract}
  Previoisly, we reported calculations of the high-temperature series for
  thermodynamical susceptibilities towards a number of possible ordered states
  in the $t$-$J$-$V$ model.  Due to an error in the calculation, the series
  for $d$-wave supeconducting and extended $s$-wave superconducting orders
  were incorrect.  We give the replacement figures.  In agreement with our
  earlier findings, we still find no evidence of any strong enhancement of the
  superconducting susceptibility with decreasing temperature.  However,
  because different Pade approximants diverge from each other at somewhat
  higher temperatures than we originally found, it is less clear what this
  implies concerning the presence or absence of high-temperature
  superconductivity in the $t$-$J$ model.
\end{abstract}

\pacs{
74.72.-h,
74.25.Dw,
74.25.Ha
}
\maketitle

Previoisly [\CITE{PKZ-2004}], we reported calculations of the high-temperature
series for thermodynamical susceptibilities towards a number of possible
ordered states in the \tjv\ model.  Due to an error in the calculation, the
series for $d$-wave supeconducting ($d$-SC) and extended $s$-wave
superconducting ($s$-SC) orders were incorrect.  The first terms of the
correct series are $$ \chi_{s-\rm SC}^{(1)}=\chi_{d-\rm
  SC}^{(1)}=\beta{(1+x)(1-3x)\over 8\ln((1-x)/2x)}.$$ Here, $\beta=1/T$ is the
inverse temperature, and the doping $x$ is the average number of holes per
site.  The SC susceptibilities at small $x\to0$ are suppressed not
quadratically but only logarithmically in $x$.  Note also that the value
$x=1/3$ is not singular, as the zeros in the numerator and the denominator
cancel.

\begin{figure}[htbp]
  \centering
  \epsfxsize=3.1in
  \epsfbox{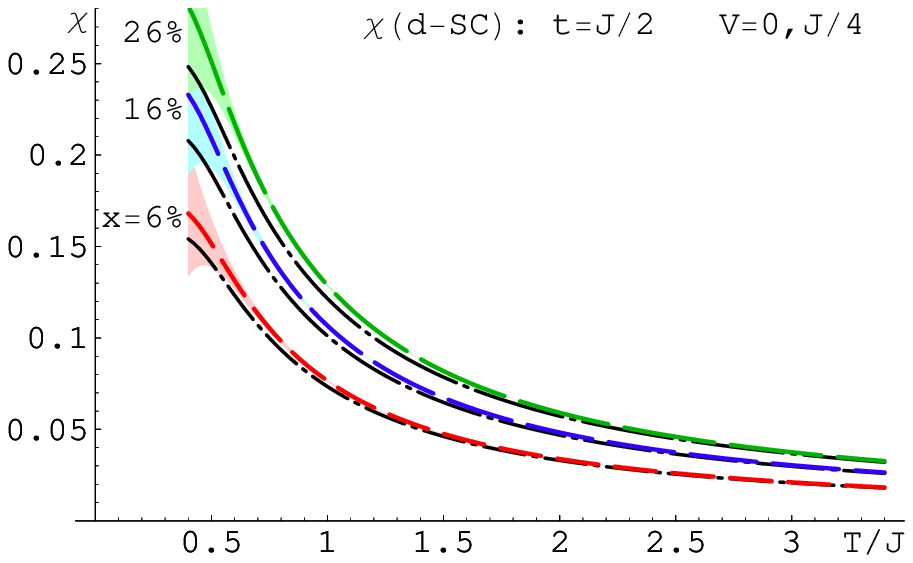}
  \caption{(Color online) Superconducting (d-wave) susceptibility for
    the ``unphysical'' case $t =J/2$ with $V=0$ (dashes) and $V=J/4$
    (dot-dash) from the HTS to $1/T^{10}$ (diagrams with up to $N_E=11$
    edges).  Shading represents the standard deviation of nondefective Pad\'e
    approximants as discussed in Ref.~\CITE{PKZ-2004}.  The corresponding
    curves for $s$-SC susceptibility (not shown) are similar but somewhat
    smaller 
    at low temperatures.}
\label{fig:err1}
\end{figure}

\begin{figure}[htbp]
  \centering
  \epsfxsize=3.1in
  \epsfbox{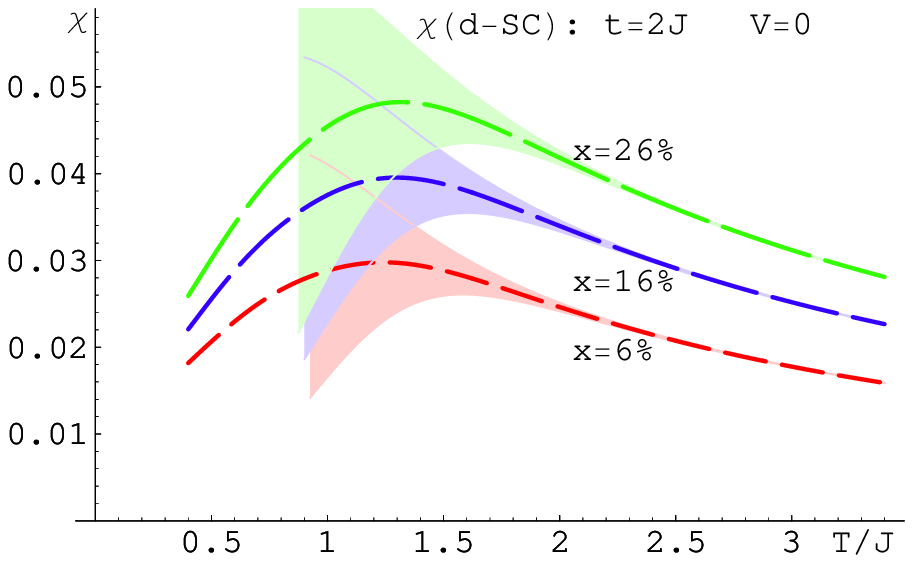}
  \caption{As in Fig.~\protect\ref{fig:err1} but for $t=2J$.  Only $V=0$
    curves are shown; curves for $V=J/4$ (not shown) are similar but more
    suppressed at low temperatures.}
\label{fig:err2}
\end{figure}

\begin{figure}[htbp]
  \centering
  \epsfxsize=3.1in
  \epsfbox{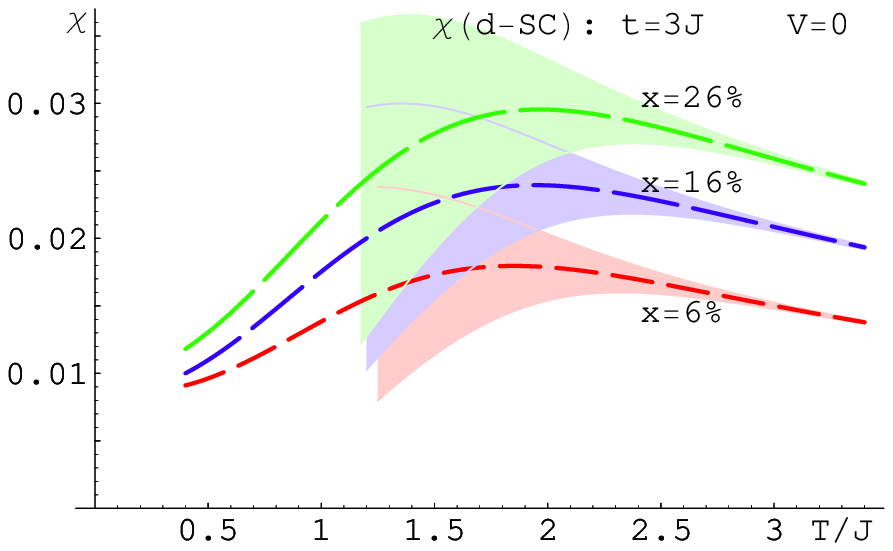}
  \caption{As in Fig.~\protect\ref{fig:err2} but for $t=3J$.   }
\label{fig:err3}
\end{figure}

We performed the corrected calculations of the series for thermodynamical
susceptibilities with an extra order in powers of $\beta$ up to the term with
$\beta^{10}$. 
The coefficients of the corrected series are available upon request.
Figs.~\ref{fig:err1} -- \ref{fig:err3} represent the analyzed results for the corrected series 
and should replace Figs.~2 and 3 in Ref.~\CITE{PKZ-2004}.

For physically plausible parameters, $t > J$, 
the tendency of the superconducting susceptibility to decrease with decreasing temperature 
is not quite as prominent as in our original work.  The increased variance
between different Pad\'es relative to our former results indicate that the
series can only be trusted down to temperatures $T\sim J$, whereas for $t=2J$,
the mean susceptibility
reaches a maximum at $T/J$ ranging from
$T/J\sim 1.2$ at $x=6\%$ to $T/J\sim 1.3$ at $x=26\%$.
Fig.~\ref{fig:err3} shows that at $t=3J$ the maximum is shifted up to
temperatures $T\sim 2J$.
We also note that the effect of the n.n.\ repulsion $V$ on the superconducting
susceptibility is 
much less dramatic than we originally reported.

As a result, although the corrected series still suggest  the absence of
high-temperature superconductivity in the \tjv\ model at $t\gtrsim 2J$, this
conclusion cannot be drawn as strongly as in our original Letter
\cite{PKZ-2004}. 

We should also note the papers by Koretsune and Ogata
\cite{koretsune-ogata-2005} and by Putikka and Luchini \cite{Putikka-2005}
which appeared after Ref.~\CITE{PKZ-2004}.  In these papers, the authors use
a similar technique to study not the thermodynamical susceptibilities but 
the \emph{equal-time} correlation functions
for the \tjv\ model with the special value $V=J/4$.  They computed the
high-temperature series for both the uniform and real-space equal-time pairing
correlation functions up to 12th order.  These authors derive an instantaneous
superconducting correlation length which increases with decreasing
temperature, but which never grows as large as one lattice constant in the
accessible range of $T/J$.  Nevertheless, they interpreted their results as an
indication that the \tj\ model has a superconducting ground state.  In our
opinion, the failure of the thermodynamical susceptibility we have computed to show any strong tendency to grow with decreasing temperature in the accessible range,
as well as the extremely short correlation lengths involved, call into
question the strength of the conclusions drawn in
Refs.~\CITE{koretsune-ogata-2005,Putikka-2005}.